\documentclass[sigconf]{acmart}

\usepackage{amssymb}

\PassOptionsToPackage{hyphens}{url}
\usepackage{url}

\usepackage{array}
\usepackage{placeins}
\usepackage{xspace}
\usepackage{loanappmacros}
\usepackage{hyphenat}
\usepackage{tabularx}
\usepackage{booktabs}
\usepackage[shortlabels]{enumitem}
\usepackage{pifont}
\usepackage{colortbl}
\usepackage{multirow}
\usepackage{xcolor}
\usepackage{subcaption}

\copyrightyear{2026}
\acmYear{2026}
\setcopyright{acmlicensed}
\acmConference[ASIA CCS '26]{ACM Asia Conference on Computer and Communications Security}{June 01--05, 2026}{Bangalore, India}
\acmBooktitle{ACM Asia Conference on Computer and Communications Security (ASIA CCS '26), June 01--05, 2026, Bangalore, India}
\acmPrice{}
\acmDOI{10.1145/3779208.3785263}
\acmISBN{979-8-4007-2356-8/2026/06}

\author{Olawale Amos Akanji}
\affiliation{%
	\institution{Boston University}
	\city{Boston}
	\state{MA}
	\country{USA}
}
\email{olawalea@bu.edu}

\author{Manuel Egele}
\affiliation{%
	\institution{Boston University}
	\city{Boston}
	\state{MA}
	\country{USA}
}
\email{megele@bu.edu}

\author{Gianluca Stringhini}
\affiliation{%
	\institution{Boston University}
	\city{Boston}
	\state{MA}
	\country{USA}
}
\email{gian@bu.edu}



\title{The Cost of Convenience: Identifying, Analyzing, and Mitigating Predatory Loan Applications on Android}

\begin{document}
	
	\begin{abstract}
		Digital lending applications, commonly referred to as \emph{loan apps}, have become a primary channel for microcredit in emerging markets. However, many of
		these apps demand excessive permissions and misuse sensitive user data for
		coercive debt-recovery practices, including harassment, blackmail, and public
		shaming that affect both borrowers and their contacts.
		
		This paper presents the first cross-country measurement of loan app compliance against both national regulations and Google's Financial Services Policy. We analyze 434 apps drawn from official registries and app markets  from Indonesia, Kenya, Nigeria, Pakistan, and the Philippines. To operationalize policy requirements at scale, we translate policy text into testable permission checks using LLM-assisted policy-to-permission mapping and combine this with static and dynamic analyses of loan apps’ code and runtime
		behavior.
		
		Our findings reveals pervasive non-compliance among \emph{approved} apps: \textbf{141} violate national regulatory policy and \textbf{147} violate Google policy. Dynamic analysis
		further shows that several apps transmit sensitive data (contacts, SMS,
		location, media) \emph{before} user signup or registration, undermining
		informed consent and enabling downstream harassment of borrowers and third
		parties. Following our disclosures, Google removed \approvedAppsremoved\
		flagged apps from Google Play, representing over 300M cumulative installs.
		
		We advocate for adopting our methodology as a proactive compliance-monitoring
		tool and offer targeted recommendations for regulators, platforms, and
		developers to strengthen privacy protections. Overall, our results highlight
		the need for coordinated enforcement and robust technical safeguards to ensure
		that digital lending supports financial inclusion without compromising user
		privacy or safety.

	\end{abstract}
	
	\maketitle
	
	\section{Introduction}
	\label{sec:intro}	
	The rapid advancement of mobile technology has revolutionized the financial services industry, fueling the expansion of Fintech products, notably digital money lending applications (\textit{loan apps}). These apps have seen explosive adoption across emerging markets like India, Indonesia, Kenya, Nigeria, Pakistan, and the Philippines. Historically, these economies relied on informal lending systems that lacked the scalability and speed needed to meet the growing credit needs of large populations, particularly those below the poverty line \cite{Flol1993MyFT, Nigeriapopulation, poverty}.
	
	Loan apps have emerged as accessible, fast alternatives to formal and informal credit systems, promising convenience and ease-of-use to millions seeking quick credit \cite{numberofloanappusers, Munyendo2022DesperateTC}. However, this convenience often comes at a steep cost: many apps obscure predatory financial terms (e.g., excessive interest rates, short repayment windows), trapping borrowers in debt cycles, and increasingly adopt unethical, digitized debt-collection tactics (harassment, blackmail) previously associated with informal lenders \cite{Aggarwal2024PredatoryLM, Munyendo2022DesperateTC}.
	
\begin{figure}[htbp]
		\centering
		\begin{minipage}[b]{0.48\columnwidth}
			\centering
			\includegraphics[width=\linewidth]{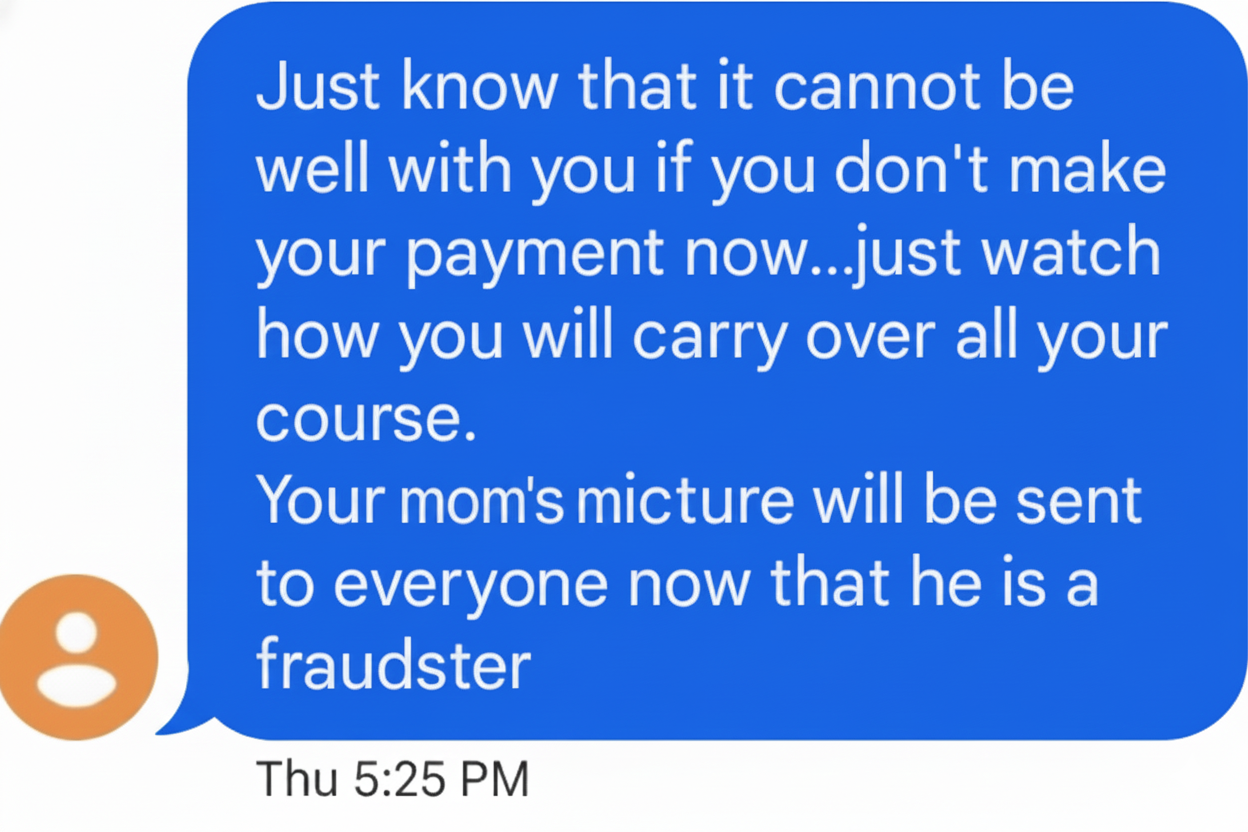}
			\subcaption{Example of a threat message received by users. \label{fig:Messagesenttouser}}
		\end{minipage}
		\hfill
		\begin{minipage}[b]{0.48\columnwidth}
			\centering
			\includegraphics[width=\linewidth]{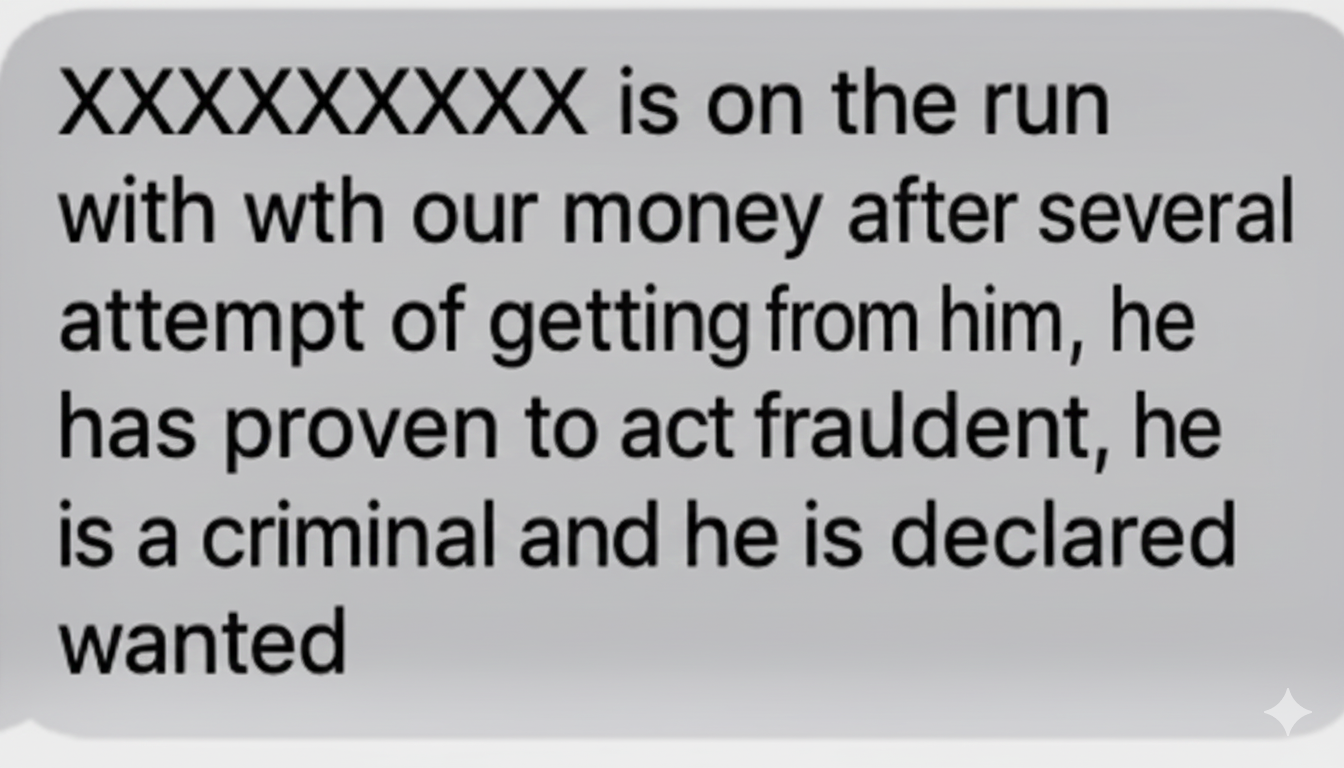}
			\subcaption{Example of a message sent to the contacts of a defaulting user. \label{fig:Messagesenttocontact}}
			
		\end{minipage}
		
		\caption{Coercive Debt Recovery Tactics}
		\Description{Two side-by-side screenshots showing coercive debt collection messages. Left: A threatening message sent directly to a loan defaulter. Right: A shaming message sent to the borrower's personal contacts to pressure repayment through social embarrassment.}
		
	\end{figure}

	Loan apps have been repeatedly linked to coercive debt collection, threats, harassment, and reputational harm that extend beyond borrowers to their families and contacts. Figures~\ref{fig:Messagesenttouser} and \ref{fig:Messagesenttocontact} illustrate examples of the threat messages and defamatory notices sent to borrowers and their contacts, highlighting the severe psychological and social impacts of these invasive practices. 
	In response to widespread abuse, governments have introduced formal
	registration and privacy constraints for digital lenders, and Google has
	introduced a Financial Services Policy (FSP) that prohibits specific
	high-risk Android permissions for lending apps~\cite{GooglePolicy}. Yet abuse
	reports persist, leading to severe psychological distress and suicides of users in
	India, Pakistan, and the Philippines~\cite{livemint_morphed,loanwebsite,
		pakistan,loanreports2,loanreports3,mexico}. Prior work on loan apps has
	primarily relied on qualitative methods, such as user interviews and case
	studies to document psychological distress and perceived data
	misuse~\cite{Munyendo2022DesperateTC,Aggarwal2024PredatoryLM,2023DEMYSTIFYINGTM,
		Ndruru2023Law,Sutedja2024An}, but does not quantitatively assess technical
	compliance with national regulations or platform policies. The persistence of
	abusive practices despite both national rules and Google’s FSP therefore
	suggests significant gaps between policy intent and enforcement reality,
	motivating our systematic, cross-country technical compliance measurement.
	
	Specifically, this study addresses the following questions:
	
	\begin{enumerate}
		\item \textbf{To what extent do loan apps targeting \emph{Indonesia, Kenya, Nigeria, Pakistan, and the Philippines} comply with national regulations and Google’s Financial Services Policy, and what misalignments or exploitable gaps exist between these policy regimes?}

		\item \textbf{How do non-compliant loan apps access, collect, and transmit sensitive user data, and what evidence confirms these data exfiltrations that may facilitate malicious activities such as threats and harassment?}
		
		\item \textbf{What regulatory measures, technical controls, and industry-wide best practices could help mitigate non-compliance, strengthen oversight, and safeguard user privacy and security in digital lending services?}
	\end{enumerate}
	
	We answer these research questions with a cross-country measurement of
	\totalApps\ Android loan apps drawn from regulator registries and app markets
	in Indonesia, Kenya, Nigeria, Pakistan, and the Philippines. Android is the
	focus of our analysis due to its dominant market share in our study regions
	(over 84\% in Africa and 83\% in Asia~\cite{statcounter_africa_os,
		statcounter_asia_os}), making it the primary platform for digital lending and
	a critical locus for regulatory enforcement. Our methodology follows the
	subcategory-specific auditing paradigm (e.g., Cardpliance and
	VioDroid-Finder~\cite{Mahmud2020CardpliancePD,Chen}), but adapts it to loan
	apps and augments it with LLM-assisted policy-to-permission mapping and
	cross-country policy analysis.
	
	Our methodology combines three components. (1) \emph{LLM-assisted
		policy-to-permission mapping} translates regulatory text into Android
	permissions, enabling automated checks against policy requirements. (2)
	\emph{Static analysis} inspects app manifests and bytecode to detect
	prohibited permissions and sensitive API calls, capturing each app’s technical
	capabilities. (3) \emph{Dynamic instrumentation and analysis} monitor runtime
	behavior to capture the timing and context of actual data access and
	transmission, distinguishing between legitimate loan or KYC workflows and
	launch-time collection that occurs before users can meaningfully consent. By
	combining these components into an end-to-end compliance pipeline, we expose
	systematic misalignment between national regulations and platform policies and
	link these gaps to concrete technical pathways for data misuse.

	In summary, this paper makes the following contributions:
	
	\begin{itemize}

		\item We present the first data‐driven measurement of loan app compliance with both national regulations (Indonesia, Kenya, Nigeria, Pakistan, and the Philippines) and Google’s Financial Services Policy. 
		
		\item We apply our methodology to \totalApps loan apps and uncover widespread non-compliance: \TotalViolateapprovedcountrypolicy approved apps violate national regulatory policies, and \TotalViolateapprovedGooglepolicy breach Google’s Financial Services Policy, with combined downloads exceeding 300 million. Following our disclosure, Google removed \approvedAppsremoved of the flagged apps. Notably, we identify 37 apps that access and transmit sensitive user data immediately upon launch, before any user interaction or registration, thereby undermining the principle of informed consent.
		
		\item We provide actionable recommendations to protect user privacy in the
		digital lending ecosystem, highlighting policy misalignments and asymmetric
		violations between national rules and Google’s Financial Services Policy. We
		recommend that Google expand its prohibited-permissions list and that
		regulators deploy automated pre-approval and periodic audits. We have shared
		our findings with Google and the relevant national regulators, and we will
		publicly release our regulatory text-to-permission mappings, LLM prompts, and
		analysis scripts to support further research, enforcement, and policy reform.
	\end{itemize}
	
	\section{Background}	
	\label{sec:background}
	This section outlines the two-tiered regulatory landscape governing loan applications: Google’s Financial Services Policy (FSP) and country-level frameworks. We detail key requirements across both regimes, including licensing mandates, specific data access restrictions, and the role of public registries in enhancing transparency. Finally, we describe the loan procurement process to identify critical stages where sensitive data collection introduces privacy risks.
	\subsection{Industry-Specific Regulation}
	Google, the primary distributor of Android apps through its Play Store, regulates digital lending through its Financial Services Policy (FSP) \cite{GooglePolicy}, which establishes explicit restrictions on data collection for personal loan apps distributed through its Play Store.

	Google’s FSP defines personal loan apps as those offering non-recurring consumer loans, excluding mortgages and revolving credit products \cite{GooglePolicy}. It requires developers to categorize such apps under the “Finance” section and provide verifiable licensing and regulatory documentation when targeting users in seven specified countries: India, Indonesia, Kenya, Nigeria, Pakistan, the Philippines, and Thailand.
	
	Beyond requiring license documentation, Google's policy strictly prohibits loan apps from requesting a defined set of eight high-risk permissions:
	{\small \texttt{READ\_EXTERNAL\_STORAGE}, \texttt{WRITE\_EXTERNAL\_STORAGE}, \texttt{READ\_MEDIA\_IMAGES}, \texttt{READ\_MEDIA\_VIDEOS}, \texttt{READ\_CONTACTS}, \texttt{READ\_PHONE\_NUMBERS}, \texttt{ACCESS\_FINE\_LOCATION}, and \texttt{QUERY\_ALL\_PACKAGES}} \cite{GooglePolicy}. The intent of these prohibitions is to protect users from harassment and cyber-bullying.
	
	\subsection{Country-Specific Regulations}
	The seven countries referenced by Google’s FSP enforce distinct national regulations designed to curb predatory lending and enhance user privacy. These frameworks generally mandate that all digital loan providers obtain proper licensing, undergo regular compliance audits, and adhere to strict limitations on data collection. Furthermore, many jurisdictions maintain public registries listing licensed and delisted providers to aid enforcement and improve user transparency.
	
	These safeguards are especially important in contexts with weak or absent central credit-reporting systems, where lenders may otherwise rely on invasive data practices to assess creditworthiness. We summarize the key regulatory provisions and prohibited data-access practices in Table~\ref{tab:loan_reg_rules_detailed}.
	
	\subsection{Loan Process and Privacy Risks}
	Loan apps typically follow a standardized process designed to deliver fast credit. This process involves installation, registration, permission granting, loan application, credit evaluation, approval, and repayment \cite{Munyendo2022DesperateTC}. While this convenience is appealing, it comes at a cost as it introduces significant privacy risks at multiple stages of the loan process.
	
	Users commonly install these apps via the Play Store, but many apps are also distributed through third-party APK sites, lender websites, or social media channels. During the initial stages, apps request access to sensitive data through two primary approaches: some demand permissions to contacts, call logs, and SMS immediately upon installation or launch, coercing users into granting access before any meaningful interaction, while others delay these requests until after registration.
	
	The privacy risks become particularly pronounced during the data collection phase. Legitimate lending only requires
	basic identifiers (e.g., phone number, government ID), yet many apps request
	access to contacts, SMS logs, call records, and device storage, often
	describing this in privacy policies as “creditworthiness assessment” even
	where regulations explicitly prohibit such collection. In cases of default,
	this surplus data is frequently weaponized for harassment, blackmail, and
	public shaming of both borrowers and their contacts, turning initial privacy
	violations into severe harm and underscoring the need for stronger,
	enforceable safeguards across platforms and regulatory systems. 
	
	In cases of default, the extensive data collection enables aggressive debt recovery tactics. Some apps misuse the collected contact information, call logs, and personal data for harassment, blackmail, and public shaming of borrowers and their contacts. These behaviors transform initial privacy violations into severe harm, highlighting the urgent need for stronger, enforceable safeguards across platforms and regulatory systems.
	\begin{figure*}[t]
		\centering
		\includegraphics[width=\textwidth, height=0.3\textheight, keepaspectratio]{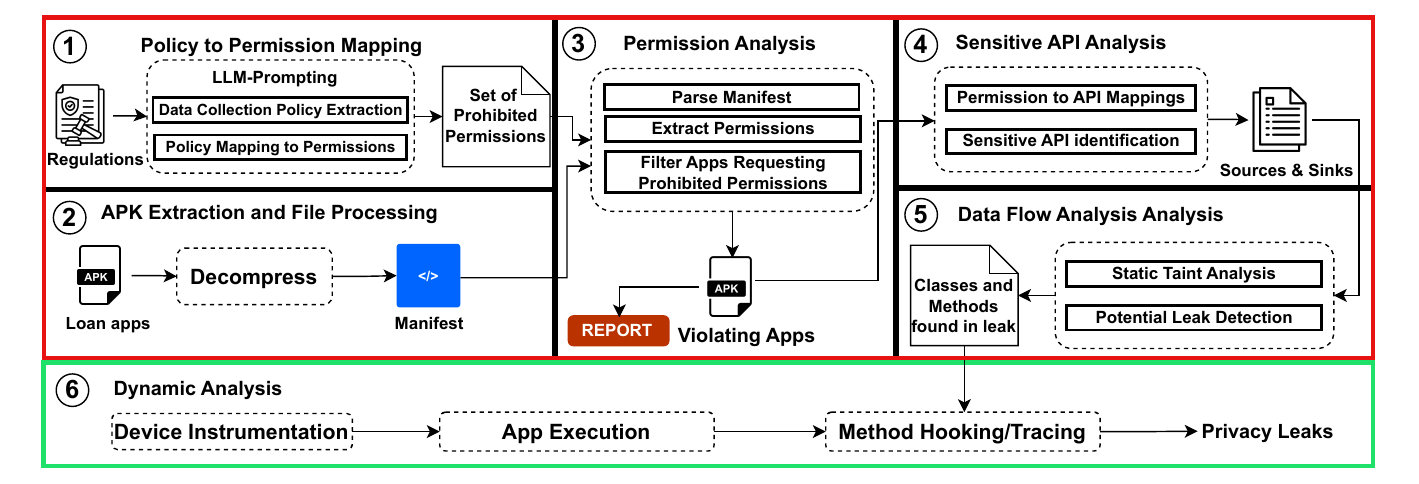}
		\caption{Overview of LoanWatch Methodology.}
		\label{fig:system}
		\Description{A flow diagram illustrating the three phases of the LoanWatch Methodology: 1) LLM-assisted policy-to-permission mapping, 2) Static Analysis of app code, and 3) Dynamic Analysis to validate runtime behavior, all leading to compliance reports.} 
	\end{figure*}
	
	\section{Methodology}
	\label{sec:methodology}
	
	To measure loan app compliance with data-collection regulations, we develop \toolname, a reproducible three-phase compliance-audit methodology (Figure~\ref{fig:system}). \toolname comprises: (1) policy extraction to map regulatory text to Android permissions, (2) static analysis to identify prohibited permissions, APIs, and potential data flows, and (3) dynamic analysis to validate runtime access and transmission of sensitive data. We consider an app \emph{violating} if it requests any permission prohibited by national regulations, Google’s Financial Services Policy (FSP), or our harmonized LoanWatch set, which combines Google’s prohibited permissions with additional high-risk permissions banned by at least one country.
	
	For policy-to-permission mapping, we use three off-the-shelf LLMs (GPT-4o Mini, Grok3, and Claude Sonnet~4) to extract prohibited data types from regulatory text, distinguish unconditional from conditional bans, and map them to Android Open Source Project (AOSP). Our static analysis inspects each app’s manifest to detect declared permissions, analyzes bytecode to identify sensitive API usage, and employs FlowDroid taint analysis to trace possible flows from sensitive sources to network sinks. Finally, semi-automated dynamic testing with Frida exercises flagged apps under guided, pre-registration interactions (launching the app and responding to permission dialogs) to confirm whether high-risk APIs actually access and transmit sensitive information.
	
	In the rest of this section, we first describe our loan app collection process, then detail each phase of \toolname.
	
	\subsection{Loan App Collection}
	\label{subsec:appcollection}
	Google's Financial Services Policy explicitly references seven countries—India, Indonesia, Kenya, Nigeria, Pakistan, the Philippines, and Thailand—where developers must provide proof of compliance with local regulations before distributing loan apps through the Play Store. However, we limit our study to five countries, Indonesia, Kenya, Nigeria, Pakistan, and the Philippines as these jurisdictions maintain publicly accessible registries. We exclude India and Thailand due to the absence of publicly available registries that list compliant or non-compliant digital lending apps at the time of data collection (July 2025).
	
	To build our dataset, we manually collect both approved and delisted apps. Approved apps are sourced directly from the Play Store to reflect their current availability and regulatory status. Delisted apps—removed for violating policy or regulatory conditions—are manually retrieved from the AndroZoo dataset and third-party APK repositories, such as APKCombo, APKMonk, and APKPure \cite{apkcombo,apkpure,apkmonk}. Including both categories enables us to compare the operational behavior of compliant versus non-compliant apps, highlighting potential differences in permission usage and privacy practices.
	
	The structure and detail of public registries varies across countries. Nigeria, Pakistan, and the Philippines maintain comprehensive registries that include both approved and delisted apps. These records typically list both company and app names, simplifying the identification of target apps. For these countries, we collect APKs by searching the Play Store, AndroZoo, and third-party repositories. In Nigeria's case, the Federal Competition and Consumer Protection Commission (FCCPC) further categorizes apps into full approval, conditional approval, watchlist, and delisted. Following consultation with the FCCPC, we collapse these into two categories for analytical clarity: "Approved" (full and conditional) and "Delisted" (watchlist and delisted).
	In contrast, Indonesia and Kenya only list approved apps; they do not maintain public records of delisted or banned apps. In Kenya, the registry lists only company names without associated apps, so we conduct manual searches to identify apps published by the listed entities. We exclude apps that are no longer operational or available on any app marketplace, as well as companies that operate exclusively via websites, since these do not request Android permissions and therefore fall outside the scope of our analysis.
	Our final dataset comprises 435 unique apps—342 approved and 93 delisted—distributed across the five selected countries.
	
	\subsection{Policy-to-Permission Mapping}
	\label{subsec:permissionmapping}
	
	Policy-to-permission mapping is essential for evaluating whether loan apps comply with regulatory data-access restrictions. Unlike Google’s Financial Services Policy, which explicitly lists banned Android permissions, national regulations often describe prohibited practices in natural language (e.g., “shall not access contacts”) without direct technical references, making automated checks non-trivial. Early privacy-policy research relied on manual expert annotation to assess compliance and policy evolution, which is accurate but cognitively costly and difficult to scale~\cite{Amos,osti_10455958}. NLP-based systems later scaled analysis through keyword detection and ML classifiers, but they require large labeled datasets and still struggle with legal nuance and domain transferability~\cite{Poligraph,arora-etal-2022-tale,Zimm}. More recent work shows that LLMs can effectively support privacy tasks such as automated data-practice identification, interactive policy assessment, and GDPR compliance evaluation~\cite{Torrado2024LargeLM,LLMpowered,morievaluating}, motivating our LLM-based approach.
	
	To establish ground truth, two Android security experts initially read each country’s regulations and manually mapped prohibited data types (contacts, call logs, SMS, media, etc.) to Android permissions. While accurate, this process required several hours per country and does not scale as regulations evolve or new jurisdictions are added. To enable reproducible and scalable audits, we therefore adopt an LLM-assisted approach that uses off-the-shelf models to generalize from legal to technical language without training task-specific classifiers.
	
	We use three LLMs—GPT-4o Mini, Grok3, and Claude Sonnet~4—to automatically map regulatory clauses to Android permissions. A structured prompting workflow (Appendix~A) defines the model persona as an Android security and data-privacy expert and provides each policy document with instructions to: (i) identify prohibited data types, (ii) distinguish unconditional from conditional prohibitions, and (iii) propose corresponding AOSP permissions. We apply this prompt to regulations from India, Indonesia, Kenya, Nigeria, the Philippines, Thailand, and Pakistan~\cite{india,indonesia,kenya,nigeriadmlguidelines,phillipinespolicydocuments,thailand,pakistanguidelines}. For each suggested permission, we verify that it exists in AOSP and that its documented capabilities match the described data access. In head-to-head comparisons against our expert-derived mappings, GPT-4o Mini and Claude Sonnet~4 cover all prohibited data types, whereas Grok3 omits several entries. We therefore construct each country’s prohibited-permission set as the union of GPT-4o Mini and Claude Sonnet~4 outputs, validated by expert review and produced in under two minutes per regulation.%
	
	For our static compliance analysis, we focus exclusively on \emph{unconditional} prohibitions—permissions that regulators ban irrespective of user consent. Conditional prohibitions depend on timing and user interactions (e.g., only during loan or KYC flows) and are therefore better evaluated, at least in part, via dynamic analysis. Restricting the static check to unconditional bans provides a clear, objective baseline for assessing compliance. This yields three prohibited-permission sets
	
	\begin{table*}[!t]
		\centering
		\small
		\caption{\textbf{LLM-Generated Country-Specific Permission Mappings}}
		\label{tab:loan_reg_rules_detailed}
		\renewcommand{\arraystretch}{1.15}
		\setlength{\tabcolsep}{2.5pt}  
		
		\begin{tabular}{|l|l|l|l|}
			\hline
			\textbf{Country} & \textbf{Category} & \textbf{Data Types} & \textbf{Mapped Android Permissions} \\
			\hline
			\textbf{India} & Unconditional & \parbox[t]{0.18\textwidth}{File \& media, Contact list, Call logs, Telephony, Biometric} & 
			\parbox[t]{0.60\textwidth}{\texttt{READ\_EXTERNAL\_STORAGE}, \texttt{WRITE\_EXTERNAL\_STORAGE}, \texttt{READ\_MEDIA\_IMAGE}, \texttt{READ\_MEDIA\_VIDEO}, \texttt{READ\_MEDIA\_AUDIO}, \texttt{MANAGE\_EXTERNAL\_STORAGE}, \texttt{READ\_CONTACTS}, \texttt{WRITE\_CONTACTS}, \texttt{GET\_ACCOUNTS}, \texttt{READ\_CALL\_LOG}, \texttt{WRITE\_CALL\_LOG}, \texttt{PROCESS\_OUTGOING\_CALLS}, \texttt{READ\_PHONE\_STATE}, \texttt{CALL\_PHONE}, \texttt{ANSWER\_PHONE\_CALLS}, \texttt{ADD\_VOICEMAIL}, \texttt{USE\_SIP}, \texttt{USE\_FINGERPRINT}, \texttt{USE\_BIOMETRIC}} \\
			\cline{2-4}
			& Conditional & \parbox[t]{0.18\textwidth}{Camera, Microphone, Location} & \parbox[t]{0.60\textwidth}{\texttt{CAMERA}, \texttt{RECORD\_AUDIO}, \texttt{ACCESS\_FINE\_LOCATION}, \texttt{ACCESS\_COARSE\_LOCATION}, \texttt{ACCESS\_BACKGROUND\_LOCATION}} \\
			\hline
			\textbf{Indonesia} & Conditional & \parbox[t]{0.18\textwidth}{All data types (with consent)} & \parbox[t]{0.60\textwidth}{All permissions (with explicit user consent)} \\
			\hline
			\textbf{Kenya} & Unconditional & \parbox[t]{0.18\textwidth}{Contact access for debt collection} & \parbox[t]{0.60\textwidth}{\texttt{READ\_CONTACTS}, \texttt{WRITE\_CONTACTS}, \texttt{GET\_ACCOUNTS}, \texttt{READ\_CALL\_LOG}, \texttt{WRITE\_CALL\_LOG}} \\
			\hline
			\textbf{Nigeria} & Unconditional & \parbox[t]{0.18\textwidth}{File \& media, Contact list, Call logs} & \parbox[t]{0.60\textwidth}{\texttt{READ\_EXTERNAL\_STORAGE}, \texttt{WRITE\_EXTERNAL\_STORAGE}, \texttt{READ\_MEDIA\_IMAGE}, \texttt{READ\_MEDIA\_VIDEO}, \texttt{READ\_MEDIA\_AUDIO}, \texttt{MANAGE\_EXTERNAL\_STORAGE}, \texttt{READ\_CONTACTS}, \texttt{WRITE\_CONTACTS}, \texttt{GET\_ACCOUNTS}, \texttt{READ\_CALL\_LOG}, \texttt{WRITE\_CALL\_LOG}, \texttt{PROCESS\_OUTGOING\_CALLS}} \\
			\hline
			\textbf{Pakistan} & Unconditional & \parbox[t]{0.18\textwidth}{Contact list, Photo gallery, SMS} & \parbox[t]{0.60\textwidth}{\texttt{READ\_CONTACTS}, \texttt{WRITE\_CONTACTS}, \texttt{GET\_ACCOUNTS}, \texttt{READ\_EXTERNAL\_STORAGE}, \texttt{WRITE\_EXTERNAL\_STORAGE}, \texttt{READ\_MEDIA\_IMAGE}, \texttt{READ\_MEDIA\_VIDEO}, \texttt{READ\_MEDIA\_AUDIO}, \texttt{MANAGE\_EXTERNAL\_STORAGE}, \texttt{READ\_SMS}, \texttt{SEND\_SMS}} \\
			\cline{2-4}
			& Conditional & Camera & \texttt{CAMERA} \\
			\hline
			\textbf{Philippines} & Unconditional & \parbox[t]{0.18\textwidth}{Contact list, Email, Social media} & \parbox[t]{0.60\textwidth}{\texttt{READ\_CONTACTS}, \texttt{WRITE\_CONTACTS}, \texttt{GET\_ACCOUNTS}} \\
			\cline{2-4}
			& Conditional & \parbox[t]{0.18\textwidth}{Camera, Photo gallery} & \parbox[t]{0.60\textwidth}{\texttt{CAMERA}, \texttt{READ\_EXTERNAL\_STORAGE}, \texttt{WRITE\_EXTERNAL\_STORAGE}, \texttt{READ\_MEDIA\_IMAGE}, \texttt{READ\_MEDIA\_VIDEO}, \texttt{READ\_MEDIA\_AUDIO}, \texttt{MANAGE\_EXTERNAL\_STORAGE}} \\
			\hline
			\textbf{Thailand} & Conditional & \parbox[t]{0.18\textwidth}{All personal data (with consent)} & \parbox[t]{0.60\textwidth}{All permissions (with explicit user consent)} \\
			\hline
		\end{tabular}
		\Description{Table showing country-specific permission mappings for seven countries (India, Indonesia, Kenya, Nigeria, Pakistan, Philippines, Thailand), categorizing data types as unconditional or conditional access, and listing corresponding Android permissions for each regulation.}
	\end{table*}

	\begin{enumerate}
		\item \textbf{Country-specific prohibited permissions sets:} Derived directly from LLM analysis of national regulatory texts, distinguishing clearly between unconditional and conditional prohibitions (Table~\ref{tab:loan_reg_rules_detailed}). Countries like Nigeria, Pakistan, and the Philippines specify explicit unconditional prohibitions, while Indonesia emphasize user consent without outright bans.
		
		\item \textbf{Google's prohibited permissions set:} As detailed in Section 2.1, Google's Financial Services Policy explicitly bans eight permissions for loan apps.
		
		\item \textbf{LoanWatch set of prohibited permissions set:} The LoanWatch set represents the union of Google's prohibited permissions and additional high-risk permissions prohibited by individual countries but not by Google. This addresses a critical gap where inconsistencies between national and platform-level prohibitions create exploitable loopholes. For example, while Google prohibits \texttt{READ\_CONTACTS}, loan apps use \texttt{READ\_CALL\_LOG} to achieve the same goal, accessing recent contact information through call history, which provides more current data about user communications for debt collection purposes. By consolidating all permissions that provide access to sensitive data types prohibited by either framework, the set prevents such circumvention tactics and ensures adequate privacy protection.
		
	\end{enumerate}

	\subsection{Static Analysis}
	\label{subsec:static_analysis}
	Static analysis is the second step in our methodology, aimed at identifying non-compliant behavior in loan apps without executing them. This phase examines each app's code and configuration to address both of our research questions: it identifies the extent of regulatory violations by analyzing declared permissions against guidelines (RQ1), and reveals the technical mechanisms through which apps can access and collect sensitive data by examining API calls and data flow paths (RQ2). By uncovering permission-based violations and mapping potential data access pathways, static analysis provides foundational evidence for understanding both compliance failures and data exploitation capabilities, setting the stage for dynamic validation.
	The static analysis pipeline consists of four core components: (i) APK Extraction and File Processing, (ii) Permission Analysis, (iii) Sensitive API Analysis, and (iv) Data Flow Analysis.
	
	\subsubsection{\textbf{APK Extraction and File Processing}}
	
	In phase 2 of our methodology, we process each app with \texttt{Androguard}~\cite{androguard}, a static-analysis framework for Android apps. Androguard unpacks the APK and exposes key artefacts for later stages: the \texttt{AndroidManifest.xml} (declared permissions and components), the \texttt{classes.dex} file (Dalvik bytecode), and related resources. We store these artefacts and use them as the input to our permission analysis, sensitive-API analysis, and data-flow analysis.
	
	\subsubsection{\textbf{Permission Analysis}}
	\label{subsubsec:permission_analysis}
	
	Android applications must explicitly declare all permissions they intend to use in their \texttt{AndroidManifest.xml}, covering both installation-time and runtime permissions~\cite{androidpermissionelement}. Using the manifests obtained in the previous step, LoanWatch parses and aggregates declared permissions for all apps in our dataset. It then compares these permissions against three reference sets: (i) the country-specific prohibited-permission set (applied only to apps targeting that country), (ii) Google’s prohibited-permission set, and (iii) the LoanWatch prohibited-permission set.
	
	We classify any app that declares at least one permission from these sets as a violating app (phase 3 of Figure~\ref{fig:system}), regardless of whether the app ultimately invokes the guarded APIs. While declaration does not prove misuse, it indicates intent and technical capacity to access sensitive data—sufficient grounds for regulatory concern when permissions are explicitly banned to prevent coercive or invasive practices. For every flagged app, we identify the target jurisdiction and compile structured reports that can be shared with the corresponding national regulator and Google.

	\subsubsection{\textbf{Sensitive API Analysis}}
	Permission analysis alone identifies apps requesting prohibited permissions but does not confirm whether these permissions are actively used. Our sensitive API analysis extends this by systematically examining each app's codebase to identify the presence of sensitive API calls that correspond to prohibited permissions. This step is essential as it verifies that apps have the technical capability to programmatically invoke restricted APIs beyond mere permission declarations.
	
	\noindent \textbf{Permission-to-API Mapping Construction:} Android's permission system changes across API levels, so accurately linking sensitive APIs to prohibited permissions requires coverage across all versions present in our dataset (API 16–36). No single source provides this, so we combine multiple mappings: Axplorer for API 16–25~\cite{axplorer}, NatiDroid for API 26–29~\cite{natidroid}, and Barzolevskaia et al.\ for API 30–33~\cite{barzolevskaia2023measuring}. For API 34–36, we perform our own static analysis using Soot, generating call graphs and control-flow analyses over AOSP framework JARs to identify permission checks in newer system APIs.

	\noindent \textbf{API Usage Detection:} With comprehensive permission-to-API mappings established, we next identify which sensitive APIs are actually invoked within each loan app's code. We apply Androguard \cite{androguard} to analyze each app's \texttt{classes.dex} file, examining the Dalvik bytecode for calls to the sensitive APIs identified in our mapping phase. We record each matched API call along with its corresponding prohibited permission, noting the exact method and class context. Since we analyze legitimate loan applications rather than malicious software, we expect minimal sophisticated obfuscation techniques that would deliberately hide API usage patterns.
	This process correlates sensitive API calls with declared permissions to build an evidence-based profile of each app's technical capability to access sensitive user data.
	
	\subsubsection{\textbf{Data Flow Analysis}}
	\label{subsubsec:dataflow_analysis}
	We perform static data flow analysis to assess whether loan apps can exfiltrate sensitive user data, such as contacts, call logs, or media. While permission and API analysis confirms technical capability, it does not establish actual data transfer paths. To bridge this gap, we employ FlowDroid~\cite{flowdroid}, a static taint-analysis tool widely used for static data flow detection~\cite{slavinPVDetectorDetectorPrivacypolicy2016,hush,schindlerPrivacyLeakIdentification2022}.
	
	We configure FlowDroid to detect data flows from sensitive APIs (sources) identified during permission analysis, such as \texttt{ContactsContract.\allowbreak Contacts}, \texttt{CallLog.\allowbreak Calls}, and \texttt{MediaStore}, to network-related methods (sinks) typically used for data transmission. Our sink definitions include FlowDroid’s default set (\texttt{HttpURLConnection.\allowbreak getInputStream}, \texttt{java.net.\allowbreak URLConnection}, \texttt{org.apache.http.\allowbreak \allowbreak HttpResponse.\allowbreak getEntity})~\cite{hush}, and popular third-party libraries (\texttt{OkHttp}, \texttt{Retrofit}, \texttt{OkGo}). This configuration allows us to identify apps with plausible static data flows from sensitive data access to network endpoints, highlighting potential data exfiltration (phase 5 in Figure~\ref{fig:system}).
	
	While FlowDroid may produce false positives due to over-approximation or struggle with obfuscation~\cite{hush, androidleaks}, we mitigate these limitations by focusing our analysis on APIs associated with declared permissions. This targeted approach reduces noise and flags apps demonstrating realistic technical pathways for sensitive data exfiltration, thereby highlighting genuine operational privacy risks.
	\subsection{Dynamic Analysis}
	\label{subsec:dynamic_analysis}
	Dynamic analysis validates the runtime behavior of loan apps by directly observing how they access, collect, and transmit sensitive user data during execution~\cite{androidleaks,Enck,Andrubis,recon}. While static analysis identifies declared permissions and potential data-flow paths, dynamic analysis provides concrete evidence of whether sensitive APIs are invoked in practice, under which conditions, and at what \emph{time} in the user journey. This temporal dimension is critical for regulations that permit certain data accesses only in specific contexts (e.g., during loan applications or Know Your Customer (KYC) verification). In this work, we apply dynamic analysis to (i) apps that static analysis flags as high risk (e.g., due to prohibited permissions or suspicious data-flow paths), and (ii) all apps from consent-focused jurisdictions such as Indonesia, where regulations emphasize timing and consent rather than outright bans. Our structured workflow comprises two primary stages: (i) device setup and instrumentation, establishing controlled runtime monitoring conditions; and (ii) app execution and method tracing, systematically capturing detailed evidence of sensitive data interactions, transmission paths, and remote storage endpoints.
	
	Core lending workflows often require verifiable personal information such as government-issued IDs, locally registered phone numbers, selfies, or bank-account details tied to domestic infrastructure. Within the scope of this study, it is neither feasible nor appropriate to supply such information or fabricate borrower identities at scale, and many apps additionally enforce geo-blocking, SIM-based checks, or emulator-detection defenses. As a result, we cannot systematically drive apps through full loan and KYC workflows.
	
	Consequently, we explicitly focus our dynamic analysis on \emph{pre-registration} behaviors. Several loan apps request sensitive permissions immediately upon launch, forcing users to grant access before they can explore the app or understand its services. This coercive approach enables apps to collect sensitive user data even when users subsequently abandon the loan application process. Our analysis examines this pattern, focusing on whether apps access contacts, call logs, and other sensitive data immediately after initial permission grants, regardless of whether users complete the loan workflow.
	
	\subsubsection{\textbf{Device Setup and Instrumentation}}
	
	To simulate typical user environments, we conduct dynamic analysis on a
	rooted Huawei Nexus~6P, an affordable Android device representative of those
	commonly used by low-income individuals in our study regions. Root access
	allows us to bypass OS-level restrictions and enables low-level
	instrumentation for behavioral monitoring. Using Android Debug Bridge (ADB)~\cite{androidadb},
	we install each loan app and prepare it for controlled execution.
	
	For runtime inspection, we use Frida~\cite{frida}, a dynamic instrumentation
	framework that injects lightweight stubs into the app's runtime. These stubs
	enable external hooks to monitor method calls, allowing us to inspect
	arguments, return values, and object states without modifying the original
	app logic. We generate a custom Frida script for each app based on the
	sensitive data-flow paths identified during static analysis. These scripts
	target methods associated with sensitive sources and sinks—such as
	\texttt{ContactsContract.Contacts.CONTENT\_URI} and
	\texttt{OkHttpClient.newCall()}—and are designed to log relevant contextual
	information, including accessed data and transmission endpoints. This
	automation ensures consistency across apps, with only minimal manual
	verification required to confirm proper hook attachment and logging
	functionality\footnote{Our regulatory text-to-permission mappings, prompting templates, and analysis scripts are available at
		\url{https://github.com/marshalwahlexyz1/-The-Cost-of-Convenience-Identifying-Analyzing-and-Mitigating-Predatory-Loan-Applications-on-Android}. \cite{OlawaleLoanAppCode2025} }.
	
	\subsubsection{\textbf{App Execution and Method Tracing}}
	
	Following setup, we execute each instrumented app and monitor whether the
	sensitive data-flow paths flagged during static taint analysis are activated
	at runtime under pre-registration conditions. Our objective is to validate
	whether sensitive permissions declared in the manifest are actually exercised
	by confirming that associated APIs are invoked and used to transmit user
	data. Frida's instrumentation logs both input (e.g., contact records, SMS
	content) and output (e.g., HTTP requests, endpoints) in real time, offering
	direct evidence of how the app handles sensitive data after permissions are
	granted.
	
	A key limitation arises from Android's on-demand class loading. Unlike
	systems that load the full codebase at startup, Android loads classes only
	when invoked. This means that some sensitive API methods may not be loaded
	during our pre-registration analysis phase, as they are triggered only by
	specific user actions or authentication flows that occur later in the app
	lifecycle. To address this limitation, we explicitly interpret our findings
	as characterizing sensitive data access during the initial app execution
	phase—specifically, APIs that are invoked immediately when apps request and
	obtain sensitive permissions upon launch.
	
	While this approach may miss apps that call sensitive APIs only in
	later-loaded classes, it effectively captures those with coercive
	permission-requesting behavior that immediately access and transmit sensitive
	data upon permission grant. For these apps, we obtain concrete runtime
	evidence of exploitation during the pre-registration phase, showing that
	users’ data can be exfiltrated even if they never complete a loan
	application.

	\section{Results}
	\label{sec:results}
	In this section, we present the results of our cross-country measurement of
	loan-app compliance with regulatory guidelines across Indonesia, Kenya,
	Nigeria, Pakistan, and the Philippines. We apply \toolname\ to a dataset of
	435 loan apps, including both approved and delisted apps from public
	registries, to identify privacy violations and validate runtime behaviors.
	Together, the static and dynamic analyses address our research questions on
	the extent of regulatory violations and the mechanisms of unauthorized data
	access, exposing enforcement gaps and informing stricter oversight. 
	
	\subsection{Static Analysis Results}
	We first present the results of our static analysis phase: permission analysis,
	sensitive-API analysis, and data-flow analysis. Together, these components use
	manifests, bytecode, and source-to-sink paths to quantify compliance
	violations, characterize policy mismatches, assess the impact of our
	harmonized prohibited-permission set, and identify apps technically capable of
	exfiltrating sensitive data without informed consent.

	\subsubsection{\textbf{Widespread Non-Compliance}}
	
	Table~\ref{tab:violations_numerical} summarizes violations for apps listed as
	Approved (available on Google Play at analysis time) and Delisted in each
	country's registry. Across all 435 loan apps analyzed, \textbf{188 (43.2\%)}
	violate at least one \emph{country} policy, and \textbf{191 (43.9\%)}
	violate Google's Financial Services Policy (FSP). Among these, 141 of the 188
	apps violating country policy are approved apps, and 147 of the 191 apps
	violating Google’s FSP are approved apps.
	
	Violations among delisted apps are unsurprising, as they may have been removed
	due to non-compliance. The high rate of violations among \emph{approved} apps
	is more concerning. Despite a dual-layered regulatory framework requiring
	compliance with both Google’s FSP and country-specific regulations, many
	approved apps still request sensitive permissions explicitly prohibited under
	these policies. This widespread non-compliance reveals critical enforcement
	gaps, allowing non-compliant apps to remain on Google Play and exposing
	borrowers to persistent privacy risks and predatory practices such as
	harassment and blackmail.
	
	\begin{table*}[t]
		\centering
		\footnotesize
		\caption{Summary of Loan App Violations by Country}
		\label{tab:violations_numerical}
		\begin{tabular}{l rr *{6}{c}} 
			\toprule
			& \multicolumn{2}{c}{\textbf{Regulator Registry}}
			& \multicolumn{2}{c}{\textbf{Violate - Country Policy}}
			& \multicolumn{2}{c}{\textbf{Violate - Google Policy}}
			& \multicolumn{2}{c}{\textbf{Violate - LoanWatch Harmonized}} \\
			\cmidrule(lr){2-3}\cmidrule(lr){4-5}\cmidrule(lr){6-7}\cmidrule(lr){8-9}
			\textbf{Country}
			& \textbf{Approved} & \textbf{Delisted}
			& \textbf{Approved} & \textbf{Delisted}
			& \textbf{Approved} & \textbf{Delisted}
			& \textbf{Approved} & \textbf{Delisted} \\
			\midrule
			Indonesia& 52 & 0 & 0/52 (0\%) & 0/0 (N/A) & 26/52 (50.0\%) & 0/0 (N/A) & 27/52 (51.9\%) & 0/0 (N/A) \\
			Kenya & 32 & 0 & 13/32 (40.6\%) & 0/0 (N/A) & 15/32 (46.9\%) & 0/0 (N/A) & 24/32 (75.0\%) & 0/0 (N/A) \\
			Nigeria & 191 & 54 & 92/189 (48.7\%) & 35/54 (64.8\%) & 69/189 (36.5\%) & 30/54 (55.6\%) & 145/189 (76.7\%) & 49/54 (90.7\%) \\
			Pakistan & 11 & 28  & 9/11 (81.8\%) & 9/28 (32.1\%) & 9/11 (81.8\%) & 9/28 (32.1\%) & 9/11 (81.8\%)  & 9/28 (32.1\%) \\
			Philippines & 56 & 10 & 27/55 (49.1\%) & 3/11 (27.3\%) & 28/55 (50.9\%) & 5/11 (45.5\%) & 46/55 (83.6\%) & 8/11 (72.7\%) \\
			\midrule
			\textbf{Total} & \textbf{342} & \textbf{92} & 141/339 (41.6\%) & 47/92 (51.1\%) & 147/339 (43.4\%) & 44/92 (47.8\%) & 251/339 (74.0\%) & 66/92 (71.7\%) \\
			\bottomrule
		\end{tabular}
		\caption*{\scriptsize Note: Violations are presented as a fraction (Violating Apps / Total Analyzed Apps) followed by the percentage of the total pool.}
	\end{table*}
	\subsubsection{\textbf{Policy Mismatches and Implications}}
	To characterize the gaps between policy frameworks and their real-world
	impact, we focus on \emph{mismatch} apps—those that violate only one of the
	two policy frameworks—and identify the specific permissions that create these
	gaps. We refer to these mismatch cases as \emph{asymmetric} violations:
	apps that violate either a country’s rules or Google’s FSP, but not both.
	Conversely, apps that either violate both frameworks or comply with both form
	\emph{symmetric} cases, where national rules and Google’s policy are
	effectively aligned. We focus here on \emph{approved} apps to show how
	asymmetric violations allow developers to exploit inconsistencies between
	national regulations and Google’s FSP to obtain prohibited data while
	remaining compliant under one framework.
	
	Google's FSP prohibits a fixed set of eight high-risk permissions targeting
	channels linked to social harm, while several national regulations in our
	study instead enumerate and prohibit specific “social-harm” data types
	(contacts, call logs, SMS). This fundamental difference in approach creates
	two operational forms of asymmetric violations:
	
	\begin{enumerate}
		\item \textbf{Country-only violations:} Apps request permissions such as
		{\small\texttt{READ\_CALL\_LOG}} or {\small\texttt{READ\_SMS}} that violate
		national regulations but not Google’s FSP, enabling access to social graphs
		and communication content for predatory practices.
		
		\item \textbf{Google-only violations:} Apps request permissions such as
		{\small\texttt{ACCESS\_FINE\_LOCATION}} or
		{\small\texttt{READ\_EXTERNAL\_STORAGE}} that violate Google’s FSP but not
		national regulations, enabling location-based threats or unauthorized media
		access.
	\end{enumerate}
	
	Below, we detail these mismatches for each country, focusing on approved apps.
	
	\noindent\textbf{Indonesia.}
	Indonesia's regulations do not explicitly prohibit specific permissions, so no
	approved apps violate country-specific policies in our static analysis.
	However, \textbf{26 of 52 approved apps (50.0\%)} violate Google's FSP by
	requesting permissions banned by Google but not addressed by Indonesian
	regulations. The most commonly requested prohibited permissions across these
	26 apps include {\small\texttt{ACCESS\_FINE\_LOCATION}} (21),
	{\small\texttt{WRITE\_EXTERNAL\_STORAGE}} (14),
	{\small\texttt{READ\_EXTERNAL\_STORAGE}} (12),
	{\small\texttt{READ\_CONTACTS}} (4),
	{\small\texttt{QUERY\_ALL\_PACKAGES}} (1), and
	{\small\texttt{READ\_PHONE\_NUMBERS}} (1). All Indonesian violators are thus
	\emph{asymmetric} Google-only cases; the remaining 26 apps are symmetric
	(compliant under both frameworks).
	
	\noindent\textbf{Kenya.}
	Among 33 approved apps, \textbf{13 (39.4\%)} violate Kenya’s policy and
	\textbf{15 (45.5\%)} violate Google’s FSP. There are 4 country-only violators
	({\small\texttt{READ\_CALL\_LOG}} in 2, {\small\texttt{GET\_ACCOUNTS}} in 2)
	and 6 Google-only violators ({\small\texttt{READ\_EXTERNAL\_STORAGE}} in 6,
	{\small\texttt{WRITE\_EXTERNAL\_STORAGE}} in 4,
	{\small\texttt{ACCESS\_FINE\_LOCATION}} in 3,
	{\small\texttt{READ\_PHONE\_NUMBERS}} in 1). The remaining 9 violators breach
	both frameworks, and 14 apps comply with both, so Kenya exhibits 10 asymmetric
	apps (4 country-only, 6 Google-only) and 23 symmetric apps (9 violating both,
	14 violating neither).
	
	\noindent\textbf{Nigeria.}
	Among 191 approved apps, \textbf{92 (48.2\%)} violate Nigeria’s policy and
	\textbf{69 (36.1\%)} violate Google’s FSP. There are 26 country-only
	violators—mostly requesting {\small\texttt{READ\_CALL\_LOG}} (23, with 2 also
	requesting {\small\texttt{GET\_ACCOUNTS}}) and
	{\small\texttt{GET\_ACCOUNTS}} alone (3)—and 3 Google-only violators that do
	not request any Nigeria-prohibited permissions but do request permissions
	banned by Google’s FSP (all three request
	{\small\texttt{ACCESS\_FINE\_LOCATION}}). The remaining 66 violators breach
	both frameworks, and 96 apps comply with both, yielding 29 asymmetric apps
	(26 country-only, 3 Google-only) and 162 symmetric apps (66 violating both,
	96 violating neither).
	
	\noindent\textbf{Pakistan.}
	Among 11 approved apps, \textbf{9 (81.8\%)} violate both Pakistan’s policy and
	Google’s FSP, with no country-only or Google-only violators. The remaining 2
	apps comply with both frameworks. Pakistan therefore exhibits exclusively
	symmetric behavior: 9 symmetric violators and 2 symmetric compliant apps, and
	no asymmetric cases.
	
	\noindent\textbf{Philippines.}
	Among 56 approved apps, \textbf{27 (48.2\%)} violate the Philippines’ policy
	and \textbf{28 (50.0\%)} violate Google’s FSP. There is 1 country-only
	violator ({\small\texttt{GET\_ACCOUNTS}} in 1 app) and 2 Google-only
	violators—apps that do not request any Philippines-prohibited permissions but
	request one or more permissions banned by Google’s FSP. Across these 2 apps,
	the flagged permissions include {\small\texttt{ACCESS\_FINE\_LOCATION}} (1)
	and both {\small\texttt{READ\_EXTERNAL\_STORAGE}} and
	{\small\texttt{WRITE\_EXTERNAL\_STORAGE}} (1). The remaining 26 violators
	breach both frameworks, and 27 apps comply with both, so the Philippines
	exhibits 3 asymmetric apps (1 country-only, 2 Google-only) and 53 symmetric
	apps (26 violating both, 27 violating neither).
	
	\noindent\textbf{Implications of policy mismatches.}
	Asymmetric violations are common across countries, showing that mismatches
	between national regulations and Google’s FSP are systemic and allow high-risk
	apps to remain on Google Play. Symmetric violators mark points of agreement on
	unacceptable data access, while country-only and Google-only violators reveal
	each framework’s blind spots: Kenya, Nigeria, and Pakistan often ban
	“side-channel” and abuse-linked permissions such as
	{\small\texttt{READ\_CALL\_LOG}} and {\small\texttt{READ\_SMS}}, which Google
	permits, whereas Google bans permissions such as
	{\small\texttt{ACCESS\_FINE\_LOCATION}} and
	{\small\texttt{READ\_EXTERNAL\_STORAGE}}, which no country restricts. These gaps create compliance “back doors,” allowing developers to
	exploit unregulated vectors without losing abusive functionality. In some
	cases, such as Indonesia, policy gaps are predominantly one-sided, rendering
	Google the sole potential line of defense, yet our results show its enforcement is weak. 
	\subsubsection{\textbf{Importance of the LoanWatch Harmonized Permission Set}}
	Our LoanWatch harmonized set consolidates all permissions prohibited by either
	national regulations or Google’s FSP, revealing the full scope of
	non-compliance and preventing apps from bypassing restrictions via alternative
	permissions that enable the same data access.
	
	Applying this harmonized policy shows that \textbf{251/342 (73.4\%)} of
	approved apps violate the combined prohibition set, compared to 141 (41.2\%)
	under country rules alone and 147 (43.0\%) under Google’s FSP alone. This
	jump reflects two key gaps: apps (i) exploit side-channel permissions (e.g.,
	{\small\texttt{READ\_CALL\_LOG}} reconstructs contact graphs when
	{\small\texttt{READ\_CONTACTS}} is blocked), and (ii) leverage omissions where
	one framework permits what the other prohibits (e.g., Google bans
	{\small\texttt{ACCESS\_FINE\_LOCATION}} while some national policies do not).
	These findings show that harmonized, policy-driven permission mapping is
	essential to capture the full space of sensitive data access beyond what any
	single framework can see.
	
	\noindent\textbf{Enforcement outcomes.}
	We disclosed our findings to Google’s Android Security team and country
	regulators in two categories: (1) 147 apps already violating existing
	policies (Google’s FSP or national regulations), and (2) an additional 104
	apps that comply with individual policies but violate the harmonized
	LoanWatch set. Alongside the app lists, we provided Google with the
	harmonized set and the rationale for including additional permissions, such
	as the fact that {\small\texttt{READ\_CALL\_LOG}} can be used as a functional
	substitute for {\small\texttt{READ\_CONTACTS}}. Google formally acknowledged receipt, confirmed removal of 93 flagged apps
	from the Play Store, and continues to investigate remaining cases. As of the
	time of writing, Nigeria’s FCCPC has also acknowledged our submissions and
	reported ongoing coordination with Google.
	
	\subsubsection{\textbf{Apps Extracting and Exfiltrating Sensitive User Data}}
	\label{subsec:data_extraction}
	
	Beyond permission requests, our static analysis examines how non-compliant
	loan apps can exploit sensitive permissions through API usage and data-flow
	paths. We analyze 317 apps flagged by the LoanWatch benchmark—spanning both
	approved and delisted apps—and confirm that each invokes at least one
	sensitive API (e.g., \texttt{ContactsContract}, \texttt{CallLog},
	\texttt{MediaStore}, \texttt{LocationManager}) and holds the
	\texttt{INTERNET} permission alongside networking libraries such as
	\texttt{HttpURLConnection} and \texttt{OkHttp}. Static taint analysis with
	FlowDroid reveals that \totalAppsWithDataLeak\ apps contain feasible
	source-to-sink paths linking sensitive data sources to network transmission
	methods. Specifically, 59 apps have flows that can leak location data, 46
	leak storage contents, 16 exfiltrate media files, 15 expose device
	identifiers (e.g., SIM serial numbers), 14 transmit contact information, 12
	send SMS content, and 8 enumerate installed packages. Several apps exhibit
	multiple leakage channels, indicating deeply embedded exfiltration logic.
	
	These technical findings align with documented real-world abuses of borrower
	harassment, blackmail, and public shaming~\cite{Munyendo2022DesperateTC,
		Aggarwal2024PredatoryLM,whatsappscam,hdbscam} and confirm that many loan apps
	do more than \emph{declare} prohibited permissions—they embed working code
	paths to harvest and transmit private user data. By moving from simple
	permission-violation counts to concrete evidence of exploit-ready data-flow
	paths from access to exfiltration, our results underscore the need for
	stronger, technically grounded enforcement mechanisms.
	
	\subsection{Dynamic Analysis Results}
	\label{subsec:dynamic_analysisresult}
	
	To validate whether the static taint paths we identified translate into
	real-world data exfiltration, we performed dynamic analysis on the 166 apps
	flagged by FlowDroid. Of these, 148 launched successfully on a rooted test
	device; the remaining 18 crashed, typically due to emulator- or
	root-detection defenses. Using Frida-based instrumentation targeted at
	sensitive APIs and network libraries, we monitored runtime invocations and
	outbound data flows without progressing into full registration workflows,
	which would require authentic personal data and raise ethical concerns. Our
	tests therefore focus on the \emph{pre-registration} phase, from first launch
	through initial permission dialogs and basic navigation.
	
	We observe a coercive permission-requesting pattern in many apps: they request
	high-risk permissions such as \texttt{READ\_CONTACTS}, \texttt{READ\_SMS},
	\texttt{ACCESS\_FINE\_LOCATION}, or \texttt{QUERY\_ALL\_PACKAGES} at startup
	and terminate or block further use if permissions are denied—effectively
	forcing users to consent before any meaningful interaction. When these
	permissions are granted, \textbf{37 of the 148} functioning apps immediately
	begin transmitting sensitive data upon launch, before any user registration or
	loan application process.
	
	A particularly aggressive subset of six apps not only enforce these
	permission gates but also trigger exfiltration of contacts, SMS messages,
	image metadata, and installed-package lists within the launcher activity
	itself (declared via \texttt{action.MAIN} and \texttt{category.LAUNCHER}),
	directly undermining Android’s runtime-permission model and the informed-consent
	expectations embedded in both national regulations and Google’s Financial
	Services Policy. Beyond these six immediately exfiltrating apps, we observe
	additional unauthorized behaviors: 29 apps transmit location data on startup,
	and 2 apps send complete package lists without user interaction. For the
	remaining apps, we do not observe pre-registration exfiltration; however, the
	exploit-ready code paths uncovered by static analysis indicate that sensitive
	data can still be harvested later in the loan or KYC workflow.
	
	Dynamic analysis also clarifies how consent-focused regulations play out in
	practice. For Indonesia, whose guidelines emphasize user consent rather than
	enumerating hard bans, we ran dynamic analysis on all 26 approved apps that
	violate only Google’s FSP in our static checks. Under our pre-registration
	workflows, none of these apps accessed contacts, call logs, SMS, or storage
	immediately upon launch; sensitive access was only triggered deeper in the
	loan or KYC flows. Thus, while these apps clearly violate Google’s FSP, our
	analysis did not observe launch-time violations of Indonesia’s
	timing-based restrictions under the tested conditions.
	
	In summary, our analysis of 434 loan apps across five countries reveals a regulatory
	landscape that systematically fails to protect users. While 73.4\% of
	approved apps (251/342) violate our harmonized prohibition set, only 41.2\%
	violate country-specific rules and 43.0\% violate Google’s policy alone,
	showing how apps exploit gaps between frameworks to access prohibited data
	while remaining technically compliant under at least one of them.
	
	Static analysis shows that all 317 apps flagged by LoanWatch invoke sensitive
	APIs together with network-capable libraries, creating the technical
	infrastructure for exfiltration. Dynamic analysis confirms that this
	infrastructure is actively abused: 37 apps begin transmitting sensitive data
	immediately upon launch, and 6 exfiltrate multiple categories of data
	(contacts, SMS, media, installed apps) from the launcher activity itself,
	before users can meaningfully engage with the app. This systematic, early-stage
	collection of highly sensitive information enables the harassment, shaming, and
	blackmail practices documented in prior work, and highlights how reactive,
	complaint-driven enforcement leaves borrowers exposed even in the presence of
	formal regulations and platform policies.
	
	\section{Discussion}
	\label{sec:discussion}
	This section consolidates our findings to highlight core challenges in
	regulating digital lending ecosystems. We discuss how failures in a
	dual-layered oversight model and fragmented policy design enable systematic
	non-compliance, link these technical violations to documented societal harms,
	and outline recommendations for regulators, platforms, developers, and users.practices in financial 
	
	\noindent\textbf{Failures in a Dual‐Layered Oversight Model.}	
	Despite the dual-layered framework involving national regulators and Google’s
	Financial Services Policy, our analysis reveals persistent enforcement gaps.
	On the regulatory side, loan apps that request explicitly prohibited
	permissions still appear in “approved” registries. Since such violations are
	immediately visible from the \texttt{AndroidManifest.xml}, this strongly
	suggests that technical checks (e.g., automated manifest inspection) are not
	being systematically applied during licensing or renewal. The absence of
	version-specific identifiers (e.g., APK hashes) in registries further prevents
	external verification that the app reviewed by regulators matches the one
	distributed to users.

	On the platform side, Google’s FSP is intended to provide an additional
	protection layer, yet we find numerous non-compliant apps available on Play
	even when they request permissions that Google itself bans. These violations
	could be detected via straightforward static analysis, but current practice
	appears to rely heavily on developer self-disclosure and complaint-driven
	enforcement. Together, these gaps in both national and platform oversight
	allow the same harmful apps to pass through multiple checkpoints and remain
	available to borrowers despite clear technical violations.

	\noindent\textbf{Policy Fragmentation and the Case for a Harmonized Set.}
	Our policy-to-permission mapping (Table~\ref{tab:loan_reg_rules_detailed})
	shows that loan app regulation is highly fragmented. Some countries (Nigeria,
	Pakistan, Philippines) unconditionally ban permissions such as
	\texttt{READ\_CALL\_LOG} and \texttt{READ\_SMS} to prevent contact harvesting
	and harassment, while others (e.g., Indonesia, Thailand) emphasize consent and
	context rather than enumerating hard bans. Google’s FSP, in contrast,
	proscribes a fixed set of eight permissions based on its own assessment of
	harassment and abuse risk.
	
	This misalignment creates exploitable loopholes. Apps can reconstruct social
	graphs via call logs when \texttt{READ\_CONTACTS} is blocked, or leverage
	location and storage permissions to facilitate threats or reputational harm
	even when local regulations do not explicitly address these channels. Our
	LoanWatch harmonized set, which merges Google’s prohibited permissions with
	additional high-risk permissions banned by at least one national regulator,
	shows that 73.4\% of approved apps would be flagged under a consolidated
	policy, compared to about 40–43\% under either framework alone.
	
	These results motivate a harmonized, conservative baseline: Google should
	extend its FSP for loan apps to cover all permissions that national
	authorities have already identified as harassment- or abuse-linked. Under
	such a policy, any loan app requesting a permission in the consolidated set
	would fail pre-approval checks, closing common “back doors” where developers
	swap one sensitive permission for another functionally equivalent one. Because
	these checks can be implemented via standard static analysis over the
	manifest, they are technically simple yet materially strengthen protection.
	
	\noindent\textbf{Societal Impact of Technical Violations.}
	These violations harm not only borrowers but also people who never interact
	with loan apps. Prior work documents coercive debt-collection practices in
	which lenders use borrowers’ private information to send threats, shaming
	messages, and defamatory broadcasts to their social networks~\cite{Munyendo2022DesperateTC,Aggarwal2024PredatoryLM,
		loanwebsite,consumercomplaints}. Our static and dynamic analyses show the
	technical basis for this behavior: harvesting contact graphs, call histories,
	SMS content, location traces, media, and app inventories, and exfiltrating
	this data to remote servers soon after permission grant.
	
	For borrowers, launch-time permission coercion and early data exfiltration
	create an immediate power imbalance: apps request high-risk permissions,
	block use if users refuse, and begin transmitting sensitive data before users
	can review terms or understand downstream uses. For third parties, the harms
	are entirely non-consensual: contacts appearing in a borrower’s address book
	or call log may receive abusive or defamatory messages despite never
	installing the app or agreeing to its terms. These “secondary victims” face
	reputational risks and relationship strain, while at a community level such
	practices erode trust in digital financial services and can push vulnerable
	users back toward informal credit channels. By linking permission use,
	API-level behavior, and observed exfiltration patterns to these direct and
	third-party harms—including documented cases of self-harm and
	suicide~\cite{livemint_morphed,pakistan,loanreports2,loanreports3}—our
	results show that privacy violations in digital lending are not merely
	procedural non-compliance but a core enabler of coercive practices.
	
	\noindent\textbf{Recommendations for Stakeholder Groups.}
	To address our third research question about what regulatory, technical, and design interventions can mitigate non-compliance, we offer recommendations for regulators, app platforms, developers, and users.
	
	\textit{\noindent \textbf{For regulators:}} Documentation-based review for licensing purposes is not enough. Regulators should adopt scalable auditing tools—such as static permission checks and code inspection—during both initial app approval and license renewal. Public registries should include version-specific identifiers like APK hashes to support external verification. In addition, regulators should establish clear channels for user complaints and whistleblowing.
	
	\textit{\noindent \textbf{For app stores and platforms:}} Google Play's reliance on self-disclosure leaves compliance gaps. We recommend automatic permission checks at submission time, especially for permissions prohibited under Google's own Financial Services Policy. The platform should also expand its policy to include omitted high risk permissions, which are banned by several national regulators.
	
	\textit{\noindent \textbf{For developers:}} Our analysis shows that several compliant loan apps deliver core lending functionality without requesting sensitive permissions. This suggests that it is technically feasible to provide digital credit services without invasive data access. Developers targeting regulated markets may benefit from minimizing permission use and reviewing their app's data access practices particularly for permissions flagged by regulators or platform policies.
	
	\textit{\noindent \textbf{For users:}} While imperfect, public registries can help users avoid unlicensed or delisted apps. Users should report apps that request unnecessary sensitive permissions either via app store feedback or national consumer protection channels. Nigeria's FCCPC, for example, maintains a reporting email for users to flag non-compliant apps \cite{FCCPCenforcement}. Still, privacy protection should not rely solely on user vigilance; systemic enforcement is essential.

	\noindent\textbf{Limitations and Future Directions.} LoanWatch relies on public registries to identify approved and delisted loan apps. As such, our dataset is bounded by countries that maintain accessible registries—namely Nigeria, Indonesia, Kenya, Pakistan, and the Philippines. Consequently, our coverage excludes countries like India or Thailand and misses unlicensed or gray-market apps, which may operate entirely outside regulatory frameworks. While this constraint narrows scope, it aligns with our objective of evaluating apps that are officially recognized by regulators and intended for lawful distribution. Future work could expand beyond licensed apps to capture unauthorized or unlisted apps via behavioral or linguistic heuristics.
	
	Ethical and operational constraints also limit our dynamic analysis. Simulating the full loan lifecycle (registration, approval, repayment, default) requires valid and verifiable identity credentials—such as phone numbers, national IDs, or bank verification numbers—to trigger backend authentication and OTP verification. Dummy or anonymized data cannot replicate these processes, restricting our visibility into post-authentication behaviors. Future work could explore partnerships with regulators or establish controlled testing environments that ethically enable full lifecycle evaluation. Despite these limitations, our methodoogy offers a conservative but reliable lower-bound assessment of compliance and risk, flagging apps violating regulations and users privacy.
	
	\section{Related Work}
	\label{sec:related}
	
	\noindent\textbf{Android Permissions and Privacy Violation Detection.}
	The Android permission model has been widely studied for over-permissioning, covert data access, and the gap between declared permissions and actual behavior. Felt et al.\ \cite{feltAndroidperm} showed that many apps request more permissions than needed, and tools such as Stowaway \cite{feltAndroidperm} detect over-privileged apps by comparing requested permissions with API use. Khatoon et al.~\cite{Khatoon2017AndroidPS} further found that free apps often exploit permissions to harvest sensitive data under ostensibly legitimate use, while later work documents systematic collection and unauthorized sharing of contacts, call logs, and SMS, raising consent and transparency concerns \cite{chen2014information,chitkaraDoesThisApp2017,Balebako2013LittleBW}. To uncover such practices, prior work developed static, dynamic, and hybrid analysis frameworks: AndroidLeaks \cite{androidleaks}, PVDetector \cite{slavinPVDetectorDetectorPrivacypolicy2016}, and FlowDroid \cite{arztSootBasedToolchainAnalyzing2017} perform taint-based static analysis; Reardon et al.\ \cite{reardon50ways} and Sarwar et al.~\cite{Sarwar2013OnTE} reveal runtime circumvention via covert channels; and Hush \cite{hush} combines static filtering with runtime confirmation. Our work directly builds on these techniques—especially static taint analysis and dynamic instrumentation—but applies them within a regulatory compliance lens tailored to loan apps’ prohibited data flows rather than general over-permissioning.
	
	\noindent\textbf{General Privacy Compliance Tools and Methods.}
	Previous work on regulatory compliance in mobile apps has primarily focused on assessing the alignment between stated privacy practices and actual data handling behaviors. Tools like MAPS \cite{maps}, POLICHECK \cite{andow2020policy}, PTPDroid \cite{slavinPVDetectorDetectorPrivacypolicy2016}, and VioDroid-Finder \cite{Chen} use natural language processing and static analysis to detect discrepancies between app's policy claims and code behavior, particularly regarding sensitive data collection (e.g., location, contacts, identifiers). VioDroid-Finder, for instance, analyzes policy structures to identify declared data access and cross-checks this against app permissions and API calls. Beyond policy-code consistency, several studies assess compliance with external regulations like GDPR and COPPA. Nguyen et al. \cite{Nguyen} identify unauthorized third-party data sharing without user consent, violating GDPR, while Fan et al. \cite{fan2020empiricalevaluationgdprcompliance}, Xiang et al. \cite{Xiang2023PolicyChecker:}, McConkey et al. \cite{McConkey2024Runtime}, and Reyes et al. \cite{Reyes2017IsOC} evaluate violations related to consent, policy completeness, and lawful basis. These approaches provide valuable methodological foundations for our work, particularly policy-to-code mapping concept. However, existing tools focus primarily on established Western regulations (GDPR, COPPA). Our work extends this paradigm by introducing LLM-assisted policy to permission mapping to handle diverse emerging market regulations and by focusing on a specific high-risk app category rather than general compliance.

	\noindent\textbf{App-Category-Specific Privacy and Compliance Audits.}
	Recent work has increasingly recognized that different app categories face distinct regulatory landscapes and risk profiles that generic compliance audits may overlook. Prior work highlights the critical need for focused audits of specific Android app subcategories due to their unique risks, access to highly sensitive data, and stringent regulatory requirements. For instance, financial apps have been evaluated for PCI DSS compliance using static analysis and UI inference~\cite{Mahmud2020CardpliancePD}, while female mHealth apps were scrutinized for sensitive data handling post-Roe v. Wade through privacy policy reviews and usability inspections~\cite{Malki2024Privacy}. IoT companion apps have been analyzed for excessive permissions via static and network analysis~\cite{Neupane2022OnTD}, and VR applications have been examined for GDPR and PIPL compliance using static analysis and NLP to identify policy inconsistencies~\cite{Zhan2024VPVet}.
	
	Within the financial app category specifically, considerable research has documented pervasive privacy breaches and aggressive debt-collection practices. Bowers et al. \cite{bowersmobilemoney} and Cranor et al. \cite{Cranor2016ALE} highlight how many finance apps omit clear disclosures about their data usage, while Butcher and James \cite{Butcher2019MicrofinanceCF} document microfinance exploitation in Latin America. In Kenya, Munyendo et al. \cite{Munyendo2022DesperateTC} rely on semi-structured user interviews to explore privacy concerns about mobile loan apps' data-collection practices and discover that users often consent to invasive permissions despite misgivings, owing to financial necessity. A similar interview-based approach by Aggarwal et al. \cite{Aggarwal2024PredatoryLM} in India identifies predatory digital lending tactics involving psychological harassment and intimidation of borrowers. Akgul et al. \cite{Akgul2024ADO} similarly find recurring complaints of data theft by loan apps in Nigeria and Indonesia. Meanwhile, Ndruru et al. \cite{Ndruru2023Law} and Sutedja et al. \cite{Sutedja2024An} analyze Indonesian online lending cases, uncovering personal-data leaks, unauthorized access, and systemic harassment facilitated by gaps in enforcement.

	Our methodology bridges these research streams by combining category-specific focus with technical compliance measurement and established privacy violation detection techniques. We focus specifically on loan apps because they represent a uniquely critical subcategory with distinct regulatory and societal implications. Unlike other app categories, loan apps operate under dual-layered regulatory constraints—both Google's Financial Services Policy (FSP) and country-specific lending regulations—while mandatorily collecting highly sensitive, non-anonymizable user data (government IDs, bank verification numbers, contacts, SMS, media files). Most critically, permission misuse in this domain directly enables coercive debt collection practices that harm borrowers and their social networks.
	
	To address the compliance challenges specific to this domain, we present a reproducible compliance-audit methodology that combines LLM-assisted policy-to-permission mapping to automatically translate regulatory text into Android permissions for precise compliance checks. This provides the first scalable, cross-country measurement of loan apps privacy violations, confirming through code-level inspection that the privacy risks reported qualitatively in prior work are both widespread and technically traceable across multiple regulatory jurisdictions.
	
	\section{Conclusion}
	\label{sec:conclusion}
	This study exposes systematic regulatory failure in digital lending oversight. Through LoanWatch, our reproducible audit methodology, we demonstrate that most approved loan apps across five countries violate data protection standards by exploiting enforcement and policy gaps to harvest sensitive user information for harassment and coercion.
	Our findings reveal widespread non-compliance, reactive enforcement, and misaligned policies that enable loan apps to systematically collect contact lists, location data, and communication records, information they weaponize against borrowers and their families.
	We call for immediate reform through harmonized permission standards, proactive technical enforcement, and strengthened coordination between platform providers and regulators to protect users.
	
	\begin{acks}
We thank the anonymous reviewers for their comments. This work was supported by the National Science Foundation under Grants CNS-2419829 and CNS-2211576. We are grateful to the Google Android Security Team for their engagement during disclosure and mitigation efforts. We also thank the various National Regulatory bodies for their essential assistance in validating the loan app registries, regulatory policy intent, and their efforts to protect consumer privacy.
\end{acks}

\bibliographystyle{ACM-Reference-Format}

	\appendix
	\section{Prompt Template for Policy-to-Permission Mapping}
	This appendix outlines the prompts used in our automated policy analysis system to map regulatory policies to Android permissions.
	
	\subsection{System Prompt}
	The system is initialized with the following role definition:
	
	\begin{quote}
		\small\textit{You are an Android security and data privacy expert with extensive knowledge of Android permissions and regulatory policies. Your expertise includes interpreting natural language policy statements and mapping them to technical permissions on the Android platform.}
	\end{quote}
	
	\subsection{User Prompt Template}
	For each policy document, the following structured prompt is used:
	
	\noindent\textbf{Task:} Analyze the following policy document and identify data access restrictions for mobile applications.
	
	\noindent\textbf{Policy Document:} \textit{[POLICY\_TEXT\_PLACEHOLDER]}
	
	\noindent\textbf{Instructions:}
	\begin{enumerate}
		\item \textbf{Identify prohibited data types:}
		\begin{itemize}
			\item Unconditionally prohibited: Data types explicitly banned regardless of user consent
			\item Conditionally prohibited: Data types that require explicit user consent
		\end{itemize}
		
		\item \textbf{Map to Android permissions:} For each identified data type, determine the corresponding Android permissions required to access that data.
		
		\item \textbf{Reasoning process:}
		\begin{enumerate}[label=\alph*.]
			\item Extract all statements relating to data access prohibitions
			\item Classify each as unconditional or conditional based on consent requirements
			\item Map prohibited data types to specific Android permissions
			\item Explain reasoning for uncertain mappings
		\end{enumerate}
	\end{enumerate}
	
	\noindent\textbf{Output Format:} For each identified restriction, provide: Data Type (as described in policy), Prohibition Type (unconditional/conditional), Android Permissions (specific permission names), and Notes (any ambiguities or uncertainties).
	
	\subsection{Example Policy Input}
	The following excerpt from India's RBI guidelines demonstrates the input format:
	
	\begin{quote}
		\small
		\textbf{Chapter IV: Technology and Data Requirement}
		
		\textbf{12. Collection, usage and sharing of data with third parties}
		
		\textbf{i.} RE shall ensure that any collection of data by their DLA and DLA of their LSP is need-based and with prior and explicit consent of the borrower having audit trail. In any case, RE shall also ensure that DLA of RE/LSP desist from accessing mobile phone resources like file and media, contact list, call logs, telephony functions, etc. 
		
		\textit{[Additional clauses omitted for brevity...]}
	\end{quote}
	
	\subsection{Sample Model Response}
	The GPT-4 model processes the policy through a systematic extraction and classification approach:
	
	\noindent\textbf{Extracted Statements:}
	\begin{itemize}
		\item From 12.i: "DLA … desist from accessing mobile phone resources like \textbf{file and media, contact list, call logs, telephony functions}, etc."
		\item From 12.i: "A one-time access can be taken for \textbf{camera, microphone, location} … only, with explicit consent."
		\item From 13.iii: "RE shall ensure that \textbf{no biometric data is stored/collected} …"
	\end{itemize}
	
	\noindent\textbf{Classification:}
	\begin{itemize}
		\item \textbf{Unconditional:} file \& media, contacts, call logs, telephony, biometric data
		\item \textbf{Conditional:} camera, microphone, location (one-time with consent)
	\end{itemize}
	
	Table~\ref{tab:permission-mapping-final-minimal} shows the complete mapping from data types to Android permissions generated by the model.
	
	\begin{table*}[t]
		\centering
		\caption{\textbf{Data Type to Android Permission Mapping (GPT-4 Response)}}
		\label{tab:permission-mapping-final-minimal}
		\small
		\renewcommand{\arraystretch}{1.15}
		\setlength{\tabcolsep}{3pt}
		
		\begin{tabular}{|l|l|l|}
			\hline
			\textbf{Data Type} & \textbf{Prohibition Type} & \textbf{Android Permission(s)} \\
			\hline
			\parbox[t]{0.18\textwidth}{File \& media} & Unconditional & \parbox[t]{0.65\textwidth}{\texttt{READ\_EXTERNAL\_STORAGE}, \texttt{WRITE\_EXTERNAL\_STORAGE}, \texttt{READ\_MEDIA\_IMAGES}, \texttt{READ\_MEDIA\_VIDEO}, \texttt{READ\_MEDIA\_AUDIO}, \texttt{MANAGE\_EXTERNAL\_STORAGE}} \\
			\hline
			Contact list & Unconditional & \parbox[t]{0.65\textwidth}{\texttt{READ\_CONTACTS}, \texttt{WRITE\_CONTACTS}, \texttt{GET\_ACCOUNTS}} \\
			\hline
			Call logs & Unconditional & \parbox[t]{0.65\textwidth}{\texttt{READ\_CALL\_LOG}, \texttt{WRITE\_CALL\_LOG}, \texttt{PROCESS\_OUTGOING\_CALLS}} \\
			\hline
			\parbox[t]{0.18\textwidth}{Telephony functions} & Unconditional & \parbox[t]{0.65\textwidth}{\texttt{READ\_PHONE\_STATE}, \texttt{CALL\_PHONE}, \texttt{ANSWER\_PHONE\_CALLS}, \texttt{ADD\_VOICEMAIL}, \texttt{USE\_SIP}} \\
			\hline
			Biometric data & Unconditional & \parbox[t]{0.65\textwidth}{\texttt{USE\_BIOMETRIC}, \texttt{USE\_FINGERPRINT}} \\
			\hline
			Camera & \parbox[t]{0.15\textwidth}{Conditional (one-time)} & \texttt{CAMERA} \\
			\hline
			Microphone & \parbox[t]{0.15\textwidth}{Conditional (one-time)} & \texttt{RECORD\_AUDIO} \\
			\hline
			Location & \parbox[t]{0.15\textwidth}{Conditional (one-time)} & \parbox[t]{0.65\textwidth}{\texttt{ACCESS\_FINE\_LOCATION}, \texttt{ACCESS\_COARSE\_LOCATION}, \texttt{ACCESS\_BACKGROUND\_LOCATION}} \\
			\hline
		\end{tabular}
		
		\Description{Table showing the mapping from eight data types (file and media, contacts, call logs, telephony functions, biometric data, camera, microphone, and location) to their corresponding Android permissions, categorized by prohibition type (unconditional or conditional one-time access).}
	\end{table*}
	
\end{document}